# Solitons supported by spatially inhomogeneous nonlinear losses

Olga V. Borovkova,[1] Yaroslav V. Kartashov,[1,2] Victor A. Vysloukh,[1] Valery E. Lobanov[1], Boris A. Malomed,[3] and Lluis Torner[1]

[1]*ICFO- Institut de Ciencies Fotoniques, and Universitat Politecnica de Catalunya, Mediterranean Technology Park, 08860, Castelldefels (Barcelona), Spain*
[2]*Institute of Spectroscopy, Russian Academy of Sciences, Troitsk, Moscow Region, 142190, Russia*
[3]*Department of Physical Electronics, School of Electrical Engineering, Faculty of Engineering, Tel Aviv University, Tel Aviv 69978, Israel*
*Olga.Borovkova@icfo.es*

**Abstract:** We uncover that, in contrast to the common belief, stable dissipative solitons exist in media with uniform gain in the presence of nonuniform cubic losses, whose local strength grows with coordinate $\eta$ (in one dimension) faster than $|\eta|$. The spatially-inhomogeneous absorption also supports new types of solitons, that do not exist in uniform dissipative media. In particular, single-well absorption profiles give rise to spontaneous symmetry breaking of fundamental solitons in the presence of uniform focusing nonlinearity, while stable dipoles are supported by double-well absorption landscapes. Dipole solitons also feature symmetry breaking, but under defocusing nonlinearity.

## 1. Introduction

Spatial solitons exist in a variety of physical settings and nonlinear media. In particular, they appear as self-trapped light beams in optical waveguides. The standard knowledge is that self-focusing and self-defocusing nonlinearities in uniform media give rise to bright and dark solitons, respectively [1]. Breaking this rule requires suitable non-uniform media. In particular, periodic transverse modulations of the refractive index support bright gap solitons even in defocusing media [2,3]. Solitons with novel properties have been also predicted and demonstrated in optical and matter-wave systems with periodic modulations of the nonlinearity along the longitudinal or transverse directions [4-10]. Still, while nonlinear lattices support stable one-dimensional solitons in focusing media [11-22], a common belief is that the existence of bright solitons requires the presence of a linear lattice in the case of defocusing nonlinearities [2,3,23]. A linear lattice potential is necessary too for the formation of multipole solitons [24-28], as they do not exist in uniform focusing media. Nevertheless, it was shown recently that stable bright fundamental and higher-order solitons do exist in the absence of any linear potential in one- and multi-dimensional conservative media with a defocusing nonlinearity whose strength grows toward the periphery of the medium faster than $r^D$, where $r$ is the radial coordinate and $D$ the spatial dimension [29-31]. On mathematical grounds, the key factor explaining the existence of solitons in such a counterintuitive setting is the non-linearizability of the underlying equations for the decaying soliton tails (the formal linearization would predict, as usual, that bright solitons do not exist in this setting). A similar mechanism explains the existence of bright solitons in a more exotic model, combining a uniform self-defocusing nonlinearity and a spatially-dependent diffraction coefficient vanishing at $r \to \infty$ [32].

Solitons exist in dissipative media too. Dissipative solitons have drawn considerable attention in various physical systems in both one- and multi-dimensional settings. Stability is a fundamental issue for such solitons, because losses in the medium must be compensated by gain. However, a spatially uniform gain destabilizes any localized wavepacket by making the background around it unstable. Two solutions to this problem have been elaborated: The use of a nonlinear gain and higher-order stabilizing absorption, such as in the physical settings governed by the complex cubic-quintic Ginzburg-Landau equations [33-38], or various schemes with localized linear gain [39-48], as well as with a localized cubic gain, in the absence of higher-order losses [49]. However, the important question whether spatial shaping of *absorption* can be used for generation and stabilization of new types of dissipative solitons was not addressed yet.

In this work we show that, contrary to the above-mentioned belief, stable dissipative fundamental and higher-order solitons can exist in media with the *uniform gain*, provided that the gain is combined with non-uniform nonlinear absorption, whose local strength grows rapidly enough toward the periphery of the system.

## 2. The model

We describe the propagation of light in such a one-dimensional medium by the nonlinear Schrödinger/Ginzburg-Landau equation for the dimensionless field amplitude $q$:

$$i\frac{\partial q}{\partial \xi} = -\frac{1}{2}\frac{\partial^2 q}{\partial \eta^2} + \sigma_r q |q|^2 + ip_i q - i\gamma(\eta) q |q|^2, \qquad (1)$$

where $\xi$ and $\eta$ are the propagation distance and transverse coordinate, respectively, $p_i > 0$ is the strength of the uniform linear gain, and $\gamma(\eta) > 0$ describes the spatial profile of the nonlinear absorption. We consider focusing, zero, and defocusing Kerr-type nonlinearities, that correspond to $\sigma_r = -1, 0, +1$, respectively. Solitons are sought for as $q(\eta, \xi) = w(\eta)\exp(ib\xi)$, where $b$ is the propagation constant, and $w(\eta) = w_r(\eta) + iw_i(\eta)$ is a complex function satisfying the stationary equation,

$$-(b + ip_i)w = -(1/2)d^2w/d\eta^2 + [\sigma_r - i\gamma(\eta)]w|w|^2. \qquad (2)$$

This model can be realized in optics, by combining a spatially uniform gain and inhomogeneous doping with two-photon-absorbing elements. Experimental techniques allowing the creation of appropriate nonuniform dopant concentration profiles have been elaborated in another context [50]. Another possibility, proposed in Refs. [29-31], is to induce an effective spatial modulation of the nonlinearity by applying an external nonuniform field which affects the local detuning of uniformly distributed dopants.

We start with the consideration of the simplest setting where the strength of the nonlinear absorption in Eqs. (1) and (2) grows towards the periphery as

$$\gamma(\eta) = \sigma_i \exp(\alpha \eta^2), \qquad (3)$$

and $\gamma(\eta)$ has a single minimum at the center (a *single-well* absorption profile). By means of rescaling we fix $\alpha = 1$ and keep the strength of the two-photon absorption, $\sigma_i$, as a free parameter. First, to confirm the existence of localized solitons in this setting, it is relevant to mention that Eqs. (2) and (3) with $\sigma_r = 0$ and $p_i = 3/2$ admit an *exact* dipole-soliton solution,

$$w(\eta) = \sigma_i^{-1/2} \eta \exp[-(1-i)\eta^2/2], \qquad (4)$$

with $b = -3/2$. While this solution is unstable (see below), the nonlinearity-modulation profile in Eq. (2) with $\gamma(\eta) = \sigma_i \eta^2 \exp(\eta^2)$, instead of (3), gives rise, at $p_i = 1/2$, to a *stable exact* fundamental soliton solution,

$$w(\eta) = \sigma_i^{-1/2} \exp[-(1-i)\eta^2/2], \qquad (5)$$

with propagation constant $b = -1/2$. These examples demonstrate the above-mentioned crucially important peculiarity of this model, namely, the nonlinearizability of Eqs. (1) and (2) due to the growth of the nonlinear absorption at $|\eta| \to \infty$ [i.e., the asymptotic form of stationary solutions cannot be found using the formal linearization of Eq. (2)]. This property explains why the uniform linear gain does not imply the instability of localized solutions in this setting.

### 3. Numerical results

Systematic numerical solutions of Eqs. (2),(3) reveal that the single-well absorption landscape gives rise to fundamental solitons with symmetric and *asymmetric* profiles, i.e., to the effect of the spontaneous symmetry breaking (SSB), if the conservative part of cubic nonlinearity is focusing ($\sigma_r = -1$). Surprisingly, bright solitons in this setting may exist even in the case of zero or defocusing conservative nonlinearities ($\sigma_r = 0, +1$). In that case, the solitons always feature symmetric shapes.

Figure 1 shows generic examples of stationary solitons in the focusing and defocusing media with the single-well nonlinear-absorption profile, depicted by the red curves. In the case of the focusing nonlinearity, the family of asymmetric solitons bifurcates from the symmetric one with the increase of the linear gain, $p_i$, at the SSB (critical) point, $p_i = p_i^{cr1}$. The bifurcation is presented in Figs. 2(a) and 2(b), where the energy flow

$$U = \int_{-\infty}^{+\infty} |q(\eta)|^2 \, d\eta,$$

and propagation constant $b$ are shown versus the gain parameter, $p_i$.

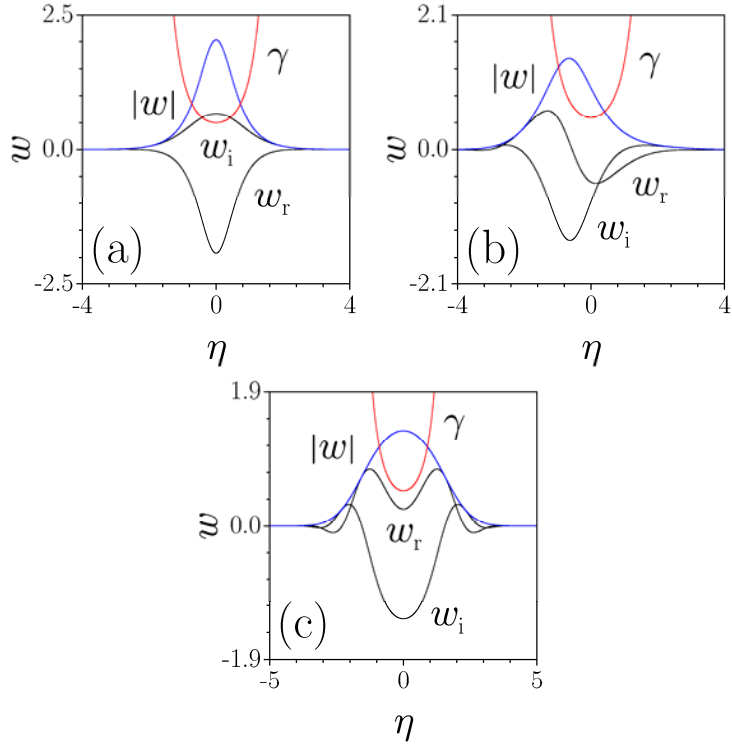

Fig. 1. (Color online) Typical profiles of fundamental symmetric (a) and asymmetric (b) solitons in the focusing medium, and of the symmetric one in the defocusing medium (c) with the single-well nonlinear-absorption profile (3) [red curves for $\gamma(\eta)$], at $p_i = 1.5$, $\sigma_i = 0.5$.

While the SSB is a common feature of double-well systems [51-54], including dissipative ones [55], it seems more surprising in the present setting, based on the single-well absorption landscape. In this connection, it is worthy to mention that a similar SSB of solitons was recently reported in a model combining the self-focusing nonlinearity, uniform cubic loss, and a localized region of linear gain [45]. Another noteworthy feature of the present setting is that, even in the case of the focusing nonlinearity, the propagation constant $b$ can be negative, while focusing nonlinearity in uniform settings gives rise to solitons with $b > 0$ (bright solitons with $b < 0$ are possible in media featuring a localized linear gain against the background of uniform linear loss [44]). Actually, symmetric solitons in Fig. 2(b) feature negative $b$ only at small values of the gain, while the asymmetric branch plunges into the region of $b < 0$ with the increase of $p_i$. The above-mentioned nonlinearizability of the underlying equation (1) is a fundamental reason for the existence of the solitons with $b < 0$ [29-31].

The dependence of the critical value of gain, at which the SSB of the fundamental solitons occurs, on the strength of the nonlinear absorption is shown in Fig. 3(a). The asymmetric solitons, which are always stable, exist above this curve, while the symmetric ones are found in the entire $(\sigma_i, p_i)$ plane, but they are stable only beneath the curve, being unstable where they coexist with the asymmetric solitons (the stability of solitons was inspected by calculating the respective eigenvalues for small perturbations). The latter property complies with the fact that the bifurcation observed in Figs. 2(a) and 2(b) is of the *supercritical* type (in other words, it is a phase transition of the second kind; note that the above-mentioned SSB bifurcation of solitons reported in Ref. [45] is also of the supercritical type).

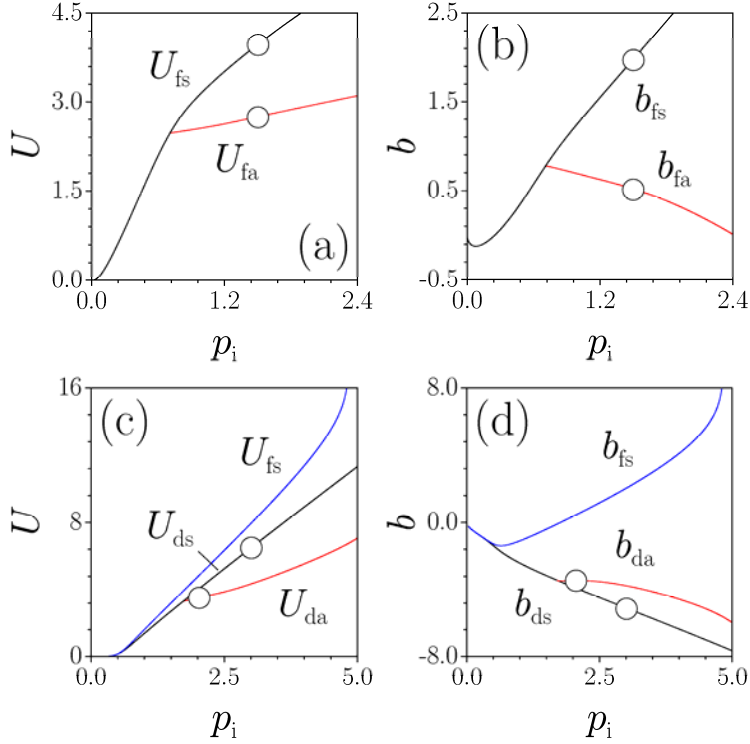

Fig. 2. (Color online) The energy flow (a) and propagation constant (b) versus $p_i$ for symmetric ("fs") and asymmetric ("fa") fundamental solitons in the focusing medium with the single-well nonlinear-absorption profile (3) at $\sigma_i = 0.5$. (c), (d): The same for dipole solitons in the double-well nonlinear-absorption landscape (6) at $\sigma_i = 2.0$. Subscripts "fs", "ds", and "da" denote symmetric dipoles in the focusing and defocusing media, and asymmetric dipoles in the defocusing medium, respectively. Circles in (a),(b) correspond to the solitons in Figs. 1(a) and 1(b), while circles in (c),(d) pertain to the solitons in Figs. 4(a) and (b), respectively.

The family of symmetric solitons that exists in the medium with the defocusing or zero conservative part of nonlinearity is *completely stable*, featuring no SSB. The energy flow of such modes increases almost linearly with the gain, while the propagation constant decreases, always staying negative.

The model with single-well absorption landscape can support higher-order (multipole) solitons too, cf. Eq. (4). This is an indication of the important fact that the spatial shaping of nonlinear losses can be used for generation of new types of solitons that do not exist in uniform dissipative systems. However, such higher-order solitons in single-well landscapes always turn out to be unstable, irrespectively of the sign of the conservative nonlinearity. The instability grows with the increase of the gain, usually transforming the multipole solitons into stable fundamental ones.

Nevertheless, *stable* multipoles (e.g., dipoles) can be found in the system with a double-well nonlinear-absorption profile, described, e.g., by function

$$\gamma(\eta) = \sigma_i (\eta^2 - \eta_0^2)^2 \exp(\alpha \eta^2), \tag{6}$$

in Eqs. (1) and (2), see examples in Fig. 4. Generic results can be adequately demonstrated for $\eta_0 = 1$. We have found that, in the focusing medium with the uniform gain and double-well profile of the nonlinear absorption, all dipole solitons are symmetric (unlike the fundamental solitons in the single-well absorption landscapes, they do not feature the SSB). The energy flow and propagation constant of the dipoles in the focusing medium are shown in Figs. 2(c) and 2(d), by curves $U_{fs}$ and $b_{fs}$, as functions of the gain. The dipoles do not exist above the upper threshold value of the gain, $p_i^{upp}$, at which tangential lines to the $U(p_i)$ and $b(p_i)$ curves become vertical [see Fig. 3(b) for domains of stability and existence of such solitons in the $(\sigma_i, p_i)$ plane]. The dipole solitons are *stable* within a limited interval of gain parameters, $p_i^{cr1} < p_i < p_i^{cr2}$ (note the difference from the symmetric fundamental solitons, which have no existence boundaries, and are completely stable in the absence of the SSB).

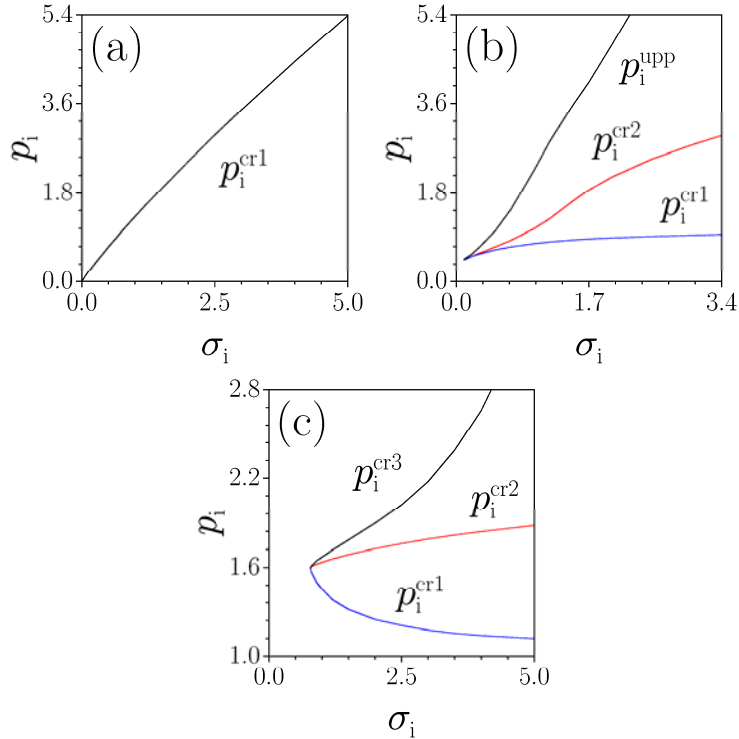

Fig. 3. (Color online) Existence and stability domains in the $(\sigma_i, p_i)$ plane. (a) The fundamental solitons in the focusing medium with the single-well nonlinear absorption profile (3) (the asymmetric solitons exist and are stable at $p_i > p_i^{cr1}$). (b) and (c): Dipole solitons in the double-well absorption profile (6), in the focusing and defocusing media, respectively.

Various scenarios of the evolution of perturbed dipoles in the focusing medium are displayed in Figs. 5(a)-5(c). At small values of the gain, $p_i < p_i^{cr1}$, the unstable dipoles usually evolve into symmetric fundamental solitons [Fig. 5(a)]. Stable propagation of the dipole soliton is demonstrated in Fig. 5(b) for $p_i^{cr1} < p_i < p_i^{cr2}$. At larger gain levels, $p_i > p_i^{cr2}$, the dipole solitons develop an oscillatory instability and may transform into persistent breathers, without symmetry breaking. The initial stage of this process is shown in Fig. 5(c), in which case the emerging breather propagates stably over an indefinitely long distance.

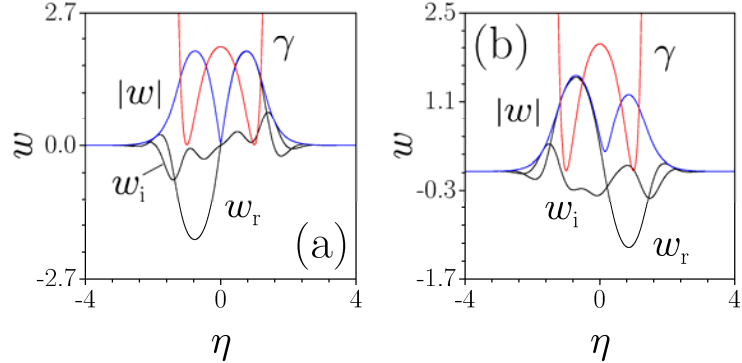

Fig. 4. (Color online) Examples of symmetric and asymmetric dipole solitons at $p_i = 3.0$ (a) and $p_i = 1.9$ (b), respectively, in the defocusing medium with the double-well nonlinear-absorption landscape (5) [red curves for $\gamma(\eta)$], at $\sigma_i = 2.0$.

On the contrary to the fundamental solitons, the dipoles exhibit the SSB in the *defocusing* medium with the double-well absorption landscape. This finding resembles a well-known fact that, in dual-core conservative systems, the SSB of spatially odd modes occurs under the action of defocusing nonlinearities (in contrast to the spatially even ground state, which undergoes the symmetry breaking under the action of focusing nonlinearities [51-54]). Examples of such symmetric and asymmetric dipoles are displayed in Fig. 4, while the symmetry-breaking bifurcation is shown by the respective $U(p_i)$ and $b(p_i)$ curves in Figs. 2(c) and 2(d). The emerging asymmetric dipoles are stable, while the symmetric ones are unstable past the bifurcation point [in compliance with the *supercritical* character of the bifurcation in Figs. 2(c) and 2(d)]. As shown in Fig. 3(c), the symmetric dipoles in the two-well absorption landscapes, with the defocusing nonlinearity, are stable in the domain of $p_i^{cr1} < p_i < p_i^{cr2}$, which also implies that they can be stable only if the nonlinear-absorption strength, $\sigma_i$, exceeds a certain minimal value. The upper stability border, $p_i = p_i^{cr2}$, is actually determined by the symmetry-breaking bifurcation. The asymmetric dipoles that emerge above the $p_i^{cr2}$ curve are also stable only within a limited range of the gain parameter, $p_i^{cr2} < p_i < p_i^{cr3}$. Figures 5(d)-5(f) illustrate the stable and unstable evolution of the dipoles in these areas. In particular, at $p_i > p_i^{cr3}$, the asymmetric dipoles are subject to an oscillatory instability, which transforms them into breathers, leading to an effective *dynamical restoration* of the symmetry [Fig. 5(f)].

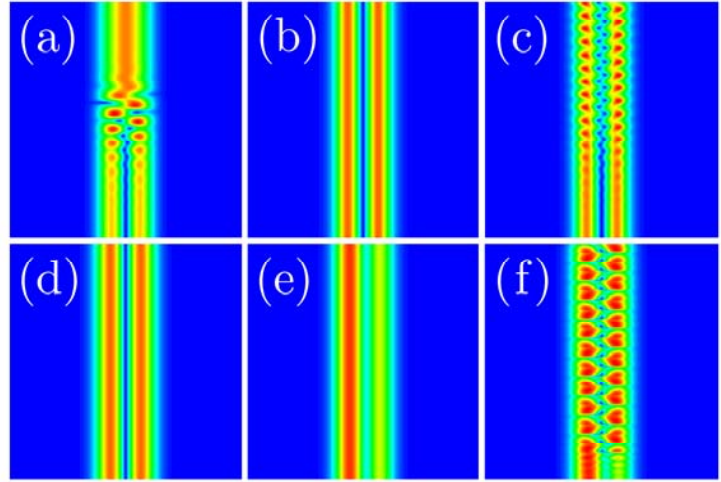

Fig. 5. (Color online) The dynamics of dipole solitons in the focusing (a)-(c) and defocusing (d)-(f) media with the double-well nonlinear-loss profile (5). (a) Spontaneous transformation of an unstable symmetric dipole into a fundamental soliton at $p_i = 0.8$. (b) Stable propagation of the symmetric dipole at $p_i = 1.8$. (c) The transformation of an unstable symmetric dipole into breather at $p_i = 3.2$. (d,e) The stable propagation of the symmetric and asymmetric dipoles at $p_i = 1.5$ and $p_i = 2.1$, respectively. (f) Spontaneous transformation of an unstable asymmetric dipole into a breather, with the dynamically restored symmetry, at $p_i = 2.23$. All the cases are shown for $\sigma_i = 3$, with small noise added to the input.

## 4. Analytical results

The numerical findings reported here can be explained in an analytical form by means of the Thomas-Fermi (TF) approximation, which, at the lowest order, neglects the diffraction term and conservative nonlinearity, i.e., all real terms, in Eq. (2) (the TF approximation was recently applied to a dissipative system with the combination of the localized linear gain and uniform linear loss in Ref. [40], but in an essentially more complex form than here). This approximation yields a real stationary wave function,

$$w_{\text{TF}}(\eta) = (p_i/\sigma_i)^{1/2} \exp(-\eta^2/2). \tag{7}$$

Strictly speaking, the TF approximation applies under the condition that $\sigma_i/\sigma_r$ and $p_i$ are large parameters, but, in reality, the accuracy of the approximation is reasonable even when these quantities take values $\sim 1$. In particular, the expression for the energy flow, $U_{\text{TF}} = \pi^{1/2} p_i/\sigma_i$, following from Eq. (7), approximates the $U(p_i)$ dependence in Fig. 2(a) quite accurately, up to the bifurcation point: for $\sigma_i = 0.5$, the slope of the linear dependence, given by this expression, is $U_{\text{TF}}/p_i \approx 3.54$, while its numerically found counterpart is $3.52$.

At the next order of the TF approximation, we substitute the wave function (7) into the real part of Eq. (2), multiply it by $w_{\text{TF}}(\eta)$, and integrate the result from $\eta = -\infty$ to $\eta = +\infty$. This procedure yields the prediction for the propagation constant, $b_{\text{TF}} = -1/4 - \sigma_r p_i/(2^{1/2} \sigma_i)$. For $\sigma_r = -1$ and $\sigma_i = 0.5$, this yields linear slope $0.71$ for the $b(p_i)$ dependence, while its numerical counterpart, corresponding to Fig. 2(b), is $0.69$.

Note that the steep anti-Gaussian modulation of the nonlinear absorption in Eq. (3) is not necessary for the existence of stable solitons in the presence of the uniform linear gain. Thus, Eq. (2) with exponential (rather than anti-Gaussian) absorption landscape, $\gamma(\eta) = a + \cosh^2(\eta)$, gives rise to an exact chirped fundamental soliton,

$$q(\eta, \xi) = (3\mu/2a)^{1/2} \exp(ib\xi) \operatorname{sech}^{1+i\mu}(\eta), \quad \mu = (3\sigma_r/2a) + \operatorname{sgn}(a)[(3\sigma_r/2a)^2 + 2]^{1/2},$$

with propagation constant $b = (1-\mu^2)/2$ and $p_i = \mu[1+(3/2a)]$. This soliton resembles the well-known Pereira-Stenflo solution [56-58], but is stable, unlike it. In this absorption landscape, an exact dipole soliton solution can be constructed too, $q(\eta, \xi) = \exp(-5i\xi/2) \operatorname{sech}^{1-i\sqrt{6}}(\eta) \sinh(\eta)$, for parameters $a = \sigma_r = 0$ and $p_i = (27/2)^{1/2}$.

Finally, using the TF approximation, it is easy to find that the mildest nonlinear-absorption profile supporting solutions with a convergent energy flow (norm) is $\gamma(\eta) \sim |\eta|^{1+\varepsilon}$ with any $\varepsilon > 0$. Furthermore, in the $D$-dimensional version of Eq. (2) the same is true for $\gamma(r) \sim |r|^{D+\varepsilon}$.

## 5. Conclusions

Our analysis has revealed that fundamental and multipole one-dimensional solitons, propagating in both focusing and defocusing nonlinear dissipative media with the uniform linear gain, can be stabilized with the help of an inhomogeneous nonlinear absorption, whose strength grows with the coordinate faster than $|\eta|$. Some soliton solutions were found in the exact analytical form. In the single-well absorption landscape, only fundamental solitons are stable, featuring the spontaneous symmetry breaking in the focusing medium. A double-well absorption landscape is required for the stabilization of dipoles, in which case the symmetry breaking occurs under the defocusing nonlinearity. Basic results for the symmetric fundamental solitons can be obtained in quite an accurate analytical form by means of the TF approximation.

It may be quite interesting to extend the analysis for two-dimensional solutions. In particular, the two-dimensional version of Eq. (2) with $\sigma_r = 0$, $p_i = 2$, and the nonlinear-loss modulation function $\sigma_i \exp(r^2)$ (here $r$ and $\theta$ are the two-dimensional polar coordinates), gives rise to an exact solution for a solitary vortex, $q = \sigma_i^{-1/2} \exp[-2i\xi + i\theta - (1-i)(r^2/2)]$, cf. Eq. (4). Further, for the modulation function $\sigma_i r^2 \exp(r^2)$ and $\sigma_r = 0$, $p_i = 1$, an exact solution for a fundamental soliton is available, $q = \sigma_i^{-1/2} \exp[-i\xi - (1-i)(r^2/2)]$, cf. Eq. (5).

## Acknowledgements

The work of O.V. Borovkova was supported by the Ministry of Science and Innovation of Spain, grant FIS2009-09928.